\documentclass[amsmath,amssymb,prb,preprintnumbers,superscriptaddress,twocolumn,showpacs,longbibliography]{revtex4-1}

\usepackage{graphicx}
\usepackage{bm}         % bold math
\usepackage{amssymb}
\usepackage{amsmath}
\usepackage{epsfig}
\usepackage[usenames]{color}
\usepackage{ulem}     % underlining and crossing of text

\newcommand{\gapprox}{{\scriptscriptstyle\stackrel{>}{\sim}}}

\begin{document}

\title{Low-Noise YBa$_2$Cu$_3$O$_7$ NanoSQUIDs for Performing Magnetization-Reversal Measurements on Magnetic Nanoparticles}

\author{T.~Schwarz}
\author{R.~W\"olbing}
\affiliation{%
Physikalisches Institut -- Experimentalphysik II and Center for Collective Quantum Phenomena in LISA$^+$,
Universit\"at T\"ubingen,
Auf der Morgenstelle 14,
D-72076 T\"ubingen, Germany}
\author{C.~F.~Reiche}
\affiliation{%
Leibniz Institute for Solid State and Materials Research IFW Dresden,
Helmholtzstr.~20,
01069 Dresden, Germany}
\author{B.~M\"uller}
\author{M.~J.~Mart\'{i}nez-P\'{e}rez}
\affiliation{%
Physikalisches Institut -- Experimentalphysik II and Center for Collective Quantum Phenomena in LISA$^+$,
Universit\"at T\"ubingen,
Auf der Morgenstelle 14,
D-72076 T\"ubingen, Germany}
\author{T.~M\"uhl}
\author{B.~B\"uchner}
\affiliation{%
Leibniz Institute for Solid State and Materials Research IFW Dresden,
Helmholtzstr.~20,
01069 Dresden, Germany}
\author{R.~Kleiner}
\author{D.~Koelle}
\affiliation{%
Physikalisches Institut -- Experimentalphysik II and Center for Collective Quantum Phenomena in LISA$^+$,
Universit\"at T\"ubingen,
Auf der Morgenstelle 14,
D-72076 T\"ubingen, Germany}

\date{\today}

\begin{abstract} %%%%%%%%%%%%%%%%%%%%%%%%%%%%%%%%%%%%%%%%%%%%%%%%%%%%%%%%%%%%%%%%%%%%%%%%%%%%%%%%%
We fabricated YBa$_2$Cu$_3$O$_7$ (YBCO) direct current (dc) nano superconducting quantum interference devices (nanoSQUIDs) based on grain boundary Josephson junctions by focused ion beam patterning.
Characterization of electric transport and noise properties at 4.2\,K in magnetically shielded environment yields a very small inductance $L$ of a few pH for an optimized device geometry.
This in turn results in very low values of flux noise $<50\,{\rm n}\Phi_0/{\rm Hz}^{1/2}$ in the thermal white noise limit, which yields spin sensitivities of a few $\mu_{\rm B}/{\rm Hz}^{1/2}$ ($\Phi_0$ is the magnetic flux quantum and $\mu_{\rm B}$ is the Bohr magneton).
We observe frequency-dependent excess noise up to 7\,MHz, which can only partially be eliminated by bias reversal readout.
This indicates the presence of fluctuators of unknown origin, possibly related to defect-induced spins in the SrTiO$_3$ substrate.
We demonstrate the potential of using YBCO nanoSQUIDs for the investigation of small spin systems, by placing a 39\,nm diameter Fe nanowire, encapsulated in a carbon nanotube, on top of a non-optimized YBCO nanoSQUID and by measuring the magnetization reversal of the Fe nanowire via the change of magnetic flux coupled to the nanoSQUID.
The measured flux signals upon magnetization reversal of the Fe nanowire are in very good agreement with estimated values, and the determined switching fields indicate magnetization reversal of the nanowire via curling mode.
\end{abstract} %%%%%%%%%%%%%%%%%%%%%%%%%%%%%%%%%%%%%%%%%%%%%%%%%%%%%%%%%%%%%%%%%%%%%%%%%%%%%%%%%


%\keywords{%Josephson junction, SQUID, nanoSQUID, spin sensitivity}
%Use showkeys class option if keyword display desired

\maketitle

\section{Introduction}
\label{sec:Introduction}

Small spin systems or magnetic nanoparticles (MNPs), like single molecular magnets, nanowires or nanotubes behave very different from magnetic bulk material, which makes them very interesting, both for basic research and applications, ranging from spintronics and spin-based quantum information processing to industrial use of ferrofluidic devices and biomedical applications \cite{Bartolome14,Bogani08,Leuenberger01,Odenbach06,Jordan99,Semelka01,Klingeler08}.
Due to their nanoscale size, MNPs have very small magnetic moments, which does not allow to use standard magnetic characterization techniques for the investigation of their properties.
In one approach, which has been pioneered by Wernsdorfer {\it et al.} \cite{Wernsdorfer01}, MNPs are placed very close to miniaturized superconducting quantum interference devices (SQUIDs), often referred to as microSQUIDs or nanoSQUIDs \cite{Awschalom88,Ketchen89,Hasselbach02,Lam03,Cleuziou06,Troeman07,Koschnick08,Hao08,Foley09,Bouchiat09,Wernsdorfer09,Giazotto10,Martinez-Perez11,Romans11,Russo12,Granata13a,Drung14}, and the magnetization reversal of MNPs is measured directly via the change of stray magnetic flux coupled to the microSQUIDs or nanoSQUIDs.
Major challenges for this application are the development of SQUIDs (i) with ultra-low flux noise, which can be achieved via the reduction of the inductance $L$ of the SQUID loop and (ii) which can be operated in very large magnetic fields (up to the Tesla range), without significant degradation of their noise performance.

The most common approach for the realization of direct current (dc) nanoSQUIDs uses two constriction-type Josephson junctions (cJJs) intersecting the SQUID loop \cite{Hasselbach02,Lam03,Troeman07,Hao08,Finkler10,Russo12,Arpaia14}.
In this case, optimum coupling between a MNP and the nanoSQUID is achieved by placing the particle directly on top of one of the cJJs.
The use of cJJs offers the possibility to operate the SQUIDs in strong magnetic fields.
However, if conventional metallic superconductors such as Pb or Nb are used, high-field operation is limited by the upper critical field of typically one Tesla for thin films \cite{Vasyukov13}.
Still, it has been demonstrated that by using ultrathin films, this limitation can be overcome \cite{Chen10a}.
However, with ultrathin films the SQUID inductance $L$ is dominated by a large kinetic inductance contribution, which yields large flux noise.
To date, the most successful approach is the SQUID-on-tip (SOT) \cite{Finkler10}.
With the so far smallest Pb SOT with 46\,nm effective loop diameter and 15\,nm film thickness, ultra-low flux noise down to $50\,{\rm n}\Phi_0/{\rm Hz}^{1/2}$ at 4.2\,K has been demonstrated \cite{Vasyukov13} ($\Phi_0$ is the magnetic flux quantum).
The inductance for a slightly larger device (56\,nm effective diameter) was estimated as $L=5.8\,$pH.
The SOT technology is extremely powerful for high-resolution scanning SQUID microscopy, and provided for the first time a spin sensitivity below $1\,\mu_\mathrm{B}/{\rm Hz}^{1/2}$ for certain intervals of applied magnetic field up to about 1 Tesla ($\mu_\mathrm{B}$ is the Bohr magneton), estimated for a point-like MNP with 10\,nm distance to the SOT.
However, maintaining the optimum flux bias point in variable magnetic field is not possible; i.e.~the flux noise and spin sensitivity strongly depend on the applied field, which makes such devices less interesting for the investigation of magnetization reversal of MNPs.

An alternative approach is the use of YBa$_2$Cu$_3$O$_7$ (YBCO) dc nanoSQUIDs with grain boundary Josephson junctions (GBJJs) for operation at temperature $T=4.2\,$K and below \cite{Nagel11}.
Magnetization reversal of a MNP can be detected by applying an in-plane magnetic field perpendicular to the grain boundary, i.e.~without significant suppression of the GBJJ critical currents.
The huge upper critical field of YBCO in the range of tens of Tesla offers the possibility for operation in strong fields up to the Tesla range, without using ultrathin films \cite{Schwarz13}.
Hence, very low inductance devices with potentially ultra-low flux noise can be realized.

Very recently, we performed an optimization study for the design of YBCO nanoSQUIDs \cite{Woelbing14}.
This is based on the calculation of the coupling factor $\phi_\mu$, i.e.~the amount of magnetic flux coupled to the SQUID per magnetic moment of a point-like MNP, placed on top of a narrow constriction inserted into the SQUID loop.
This additional constriction allows for the optimization of $\phi_\mu$ (via constriction geometry) without affecting the junctions.
In addition, we performed numerical simulations to calculate the SQUID inductance and root-mean-square (rms) spectral density of flux noise $S_\mathrm{\Phi,w}^{1/2}$ in the thermal white noise limit.
This enabled us to predict the spin sensitivity in the thermal white noise limit $S_\mathrm{\mu,w}^{1/2}=S_\mathrm{\Phi,w}^{1/2}/\phi_\mu$ for our devices as a function of all relevant device parameters.
This optimization study predicts optimum performance for a YBCO film thickness $d\approx 120\,$nm, which allows to realize nanoSQUIDs with very small $L$ of a few pH.
For optimized devices, we predict $S_\mathrm{\Phi,w}^{1/2}$ of several tens of n$\Phi_0/{\rm Hz}^{1/2}$ and $\phi_\mu\sim 10-20\,{\rm n}\Phi_0/\mu_\mathrm{B}$ (for a MNP placed 10\,nm above the YBCO film on top of the constriction), yielding a spin sensitivity $S_\mathrm{\mu,w}^{1/2}$ of a few $\mu_\mathrm{B}/{\rm Hz}^{1/2}$.

Here, we report on the realization of optimized YBCO nanoSQUIDs based on GBJJs and on the experimental determination of their electric transport and noise properties in magnetically shielded environment at $T=4.2\,$K.
To demonstrate the suitability of our YBCO nanoSQUIDs for the detection of small spin systems, we present the measurement of the magnetization reversal (up to $\sim 200\,$mT at $T=4.2\,$K) of a Fe nanowire with diameter $d_\mathrm{Fe}=39\,$nm, which was positioned close the SQUID loop.

\section{Device fabrication and experimental setup}
\label{sec:Fab-Setup}

The fabrication of the devices was carried out according to Refs.~[\onlinecite{Nagel11},\onlinecite{Schwarz13}].
A $c$-axis oriented YBCO thin film of thickness $d$ was grown epitaxially by pulsed laser deposition on a SrTiO$_3$ (STO) [001] bicrystal substrate with a 24$^\circ$ grain boundary misorientation angle.
An in-situ evaporated Au layer of thickness $d_\mathrm{Au}$ serves as shunt resistance to provide non-hysteretic current-voltage characteristics (IVCs).
SQUIDs with smallest line widths down to 50\,nm were patterned by focused ion beam (FIB) milling with 30\,keV Ga ions.
The Au layer also minimizes Ga implantation into the YBCO film during FIB milling.

For characterization of the device properties, electric transport and noise measurements were performed in electrically and magnetically shielded environment at $T=4.2\,$K, i.e.~with the samples immersed into liquid He.
By applying a modulation current $I_\mathrm{mod}$ across the constriction, the magnetic flux coupled to the SQUID can be modulated.
This allows flux biasing at the optimum working point and operation in a flux locked loop (FLL) mode \cite{Drung-SHB-4}.
In FLL mode, a deviation from the voltage at the optimum working point (due to any flux signal) is amplified and then fed back via a feedback resistor as a feedback current through the constriction.
The feedback current produces a feedback flux, cancelling the applied flux signal, i.e., the SQUID is always operated at its optimum working point, and the voltage across the feedback resistor (proportional to the flux signal) serves as the output signal.
The readout in FLL mode is limited by the bandwidth of the feedback circuit.
If the signals applied to the SQUID are small enough, one can also operate the SQUID in open loop mode, i.e., the voltage across the SQUID is amplified without feedback, and the amplified voltage serves as the output signal.
In this case, the readout is limited by the bandwidth of the voltage amplifier, which is typically larger than the FLL bandwidth.
To determine the spectral density of flux noise $S_\Phi$ vs frequency $f$ of the devices we used a Magnicon SEL-1 SQUID electronics \cite{SEL} in direct readout mode \cite{Drung03}, which was either operated in open loop mode (maximum bandwidth $\sim$7\,MHz), or in FLL mode (maximum bandwidth $\sim$500-800\,kHz).
The SEL electronics allows for SQUID operation either with constant bias current (dc bias) or with a bias reversal readout scheme (maximum bias reversal frequency $f_\mathrm{br} =260\,$kHz), to reduce $1/f$ noise caused by fluctuations of the critical currents $I_{0,1}$ and $I_{0,2}$ of the Josephson junctions 1 and 2, respectively \cite{Drung-SHB-4}.

%%%%% Fig.1 %%%%%%%%%%%%%%%%%%%%%%%%%%%%%%%%%%%%%%%%%%%%%
\begin{figure}[b]
\includegraphics[width=\columnwidth]{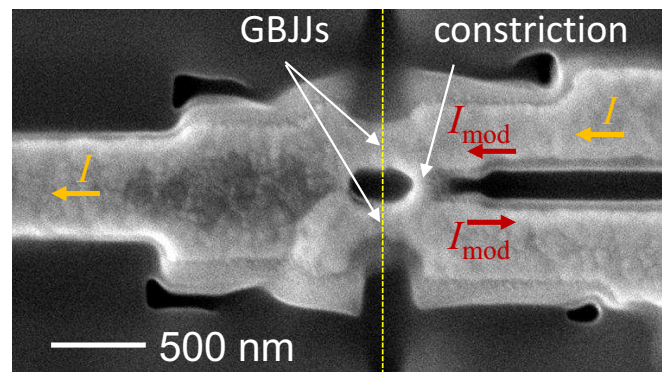}
\caption{(Color online)
SEM image of YBCO nanoSQUID-1.
Vertical dashed line indicates position of the grain boundary intersecting the two SQUID arms.
Horizontal arrows indicate paths for modulation current $I_\mathrm{mod}$ across the constriction and bias current $I$ across the grain boundary Josephson junctions.}
\label{Fig:SEM-S1}
\end{figure}
%%%%% Fig.1 %%%%%%%%%%%%%%%%%%%%%%%%%%%%%%%%%%%%%%%%%%%%%

Below we present data of our best device, SQUID-1, with a $d=120\,$nm thick YBCO film.
Figure \ref{Fig:SEM-S1} shows a scanning electron microscope (SEM) image of SQUID-1.
The loop size $350\times 190\,{\rm nm}^2$ is given by the length $l_\mathrm{J}$ of the bridges straddling the grain boundary and by the length $l_\mathrm{c}$ of the constriction.
SQUID-1 has junction widths $w_\mathrm{J1} = 210\,$nm and $w_\mathrm{J2} = 160\,$nm and a constriction width $w_\mathrm{c} = 85\,$nm.
The parameters for SQUID-1 are summarized in Table \ref{Tab:Parameters SQUID}.
For comparison, we also include parameters for a similar device, SQUID-2, which has the same YBCO film thickness, however slightly larger inductance $L=6.3\,$pH, and about a factor of 2.5 smaller characteristic voltage $V_\mathrm{c}\equiv I_\mathrm{c}R_\mathrm{N}$.
$I_\mathrm{c}$ is the maximum critical current and $R_\mathrm{N}$ is the asymptotic normal state resistance of the SQUID.
Details on electric transport and noise characteristics of SQUID-2 are presented in Sec.~I of the Supplemental Material \cite{Suppl}.
Those also include noise data taken from 6\,K to 65\,K in a different setup with a temperature stability of $\sim 1\,$mK \cite{Kemmler09}.
Table \ref{Tab:Parameters SQUID} also includes parameters for SQUID-3, which was used for measurements on an Fe nanowire in a high-field setup, as discussed further below.

\section{SQUID-1: Electric transport and noise}
\label{sec:SQUID-1}

\subsection{SQUID-1: dc characteristics}
\label{sec:SQUID-1-dc}

%%%%% Fig.2 %%%%%%%%%%%%%%%%%%%%%%%%%%%%%%%%%%%%%%%%%%%%%
\begin{figure}[t]
\includegraphics[width=\columnwidth]{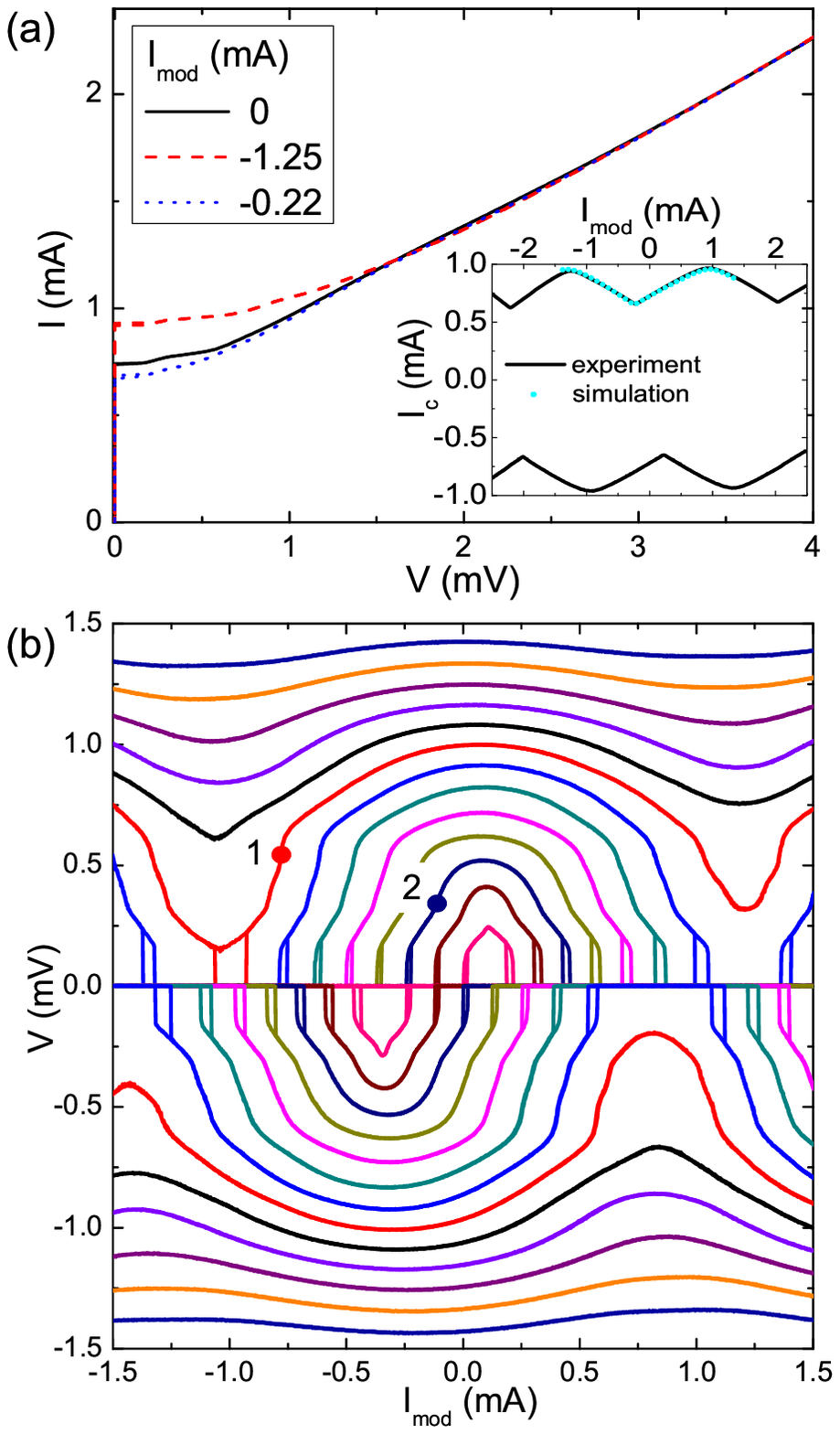}
\caption{(Color online)
SQUID-1 dc transport characteristics.
(a) Measured IVCs for three different values of $I_\mathrm{mod}$, including flux bias ($I_\mathrm{mod}$) values which yield maximum and minimum critical current.
Inset: measured $I_\mathrm{c}(I_\mathrm{mod})$ for positive and negative current bias (solid lines) and numerical simulations (dots).
(b) Measured $V(I_\mathrm{mod})$ for bias currents $|I|=0.64\ldots 1.12\,$mA (in 40\,$\mu$A steps).
Points 1 and 2 are bias points with $V_\Phi$=12 and 4.5\,mV/$\Phi_0$, respectively.}
\label{Fig:dc-S1}
\end{figure}
%%%%% Fig.2 %%%%%%%%%%%%%%%%%%%%%%%%%%%%%%%%%%%%%%%%%%%%%

Figure \ref{Fig:dc-S1} shows the dc characteristics of SQUID-1.
Figure \ref{Fig:dc-S1}(a) shows IVCs for $I_\mathrm{mod}=0$ and two values of $I_\mathrm{mod}$, corresponding to maximum and minimum critical current.
The IVCs are slightly hysteretic with maximum critical current $I_\mathrm{c}=960\,\mu$A and $R_\mathrm{N} =2.0\,\Omega$, which yields $V_\mathrm{c}= 1.92\,$mV.
The inset of Fig.~\ref{Fig:dc-S1}(a) shows the modulation of the critical current $I_\mathrm{c}(I_\mathrm{mod})$.
From the modulation period, we find for the magnetic flux $\Phi$ coupled to the SQUID by $I_\mathrm{mod}$ the mutual inductance $M=\Phi/I_\mathrm{mod}=0.44\,\Phi _0/{\rm mA}=0.91\,$pH.
We performed numerical simulations, based on the resistively and capacitively shunted junction (RCSJ) model, to solve the coupled Langevin equations which include thermal fluctuations of the junction resistances \cite{Chesca-SHB-2}.
From simulations of the $I_\mathrm{c}(I_\mathrm{mod})$ characteristics [cf.~inset of Fig.~\ref{Fig:dc-S1}(a)] we obtain for the screening parameter $\beta_L = 2I_0 L/\Phi_0 = 1.8$ (with $I_0=(I_{0,1}+I_{0,2})/2$), which yields a SQUID inductance $L =3.9\,$pH.
We do find good agreement between the measured and simulated $I_\mathrm{c}(I_\mathrm{mod})$ characteristics if we include an inductance asymmetry $\alpha _L \equiv (L_2-L_1)/(L_2+L_1)= 0.20$ ($L_1$ and $L_2$ are the inductances of the two SQUID arms) and a critical current asymmetry $\alpha_I$ $\equiv (I_{0,2}-I_{0,1})/(I_{0,2}+I_{0,1}) = 0.27$.
These asymmetries are caused by asymmetric biasing of the SQUID and by asymmetries of the device itself.

$V(I_\mathrm{mod})$ is plotted in Fig.~\ref{Fig:dc-S1}(b) for different bias currents.
The transfer function, i.e.~the maximum value of $\partial V/\partial\Phi$, in the non-hysteretic regime is $V_\Phi\approx 12\,{\rm mV}/\Phi_0$ [at $I=0.92\,$mA; cf.~point 1 in Fig.~\ref{Fig:dc-S1}(b)].

\subsection{SQUID-1: Noise data}
\label{subsec:SQUID-1-noise}

\subsubsection{Open loop mode}
\label{subsubsec:open-loop}

\begin{figure}[tb]
\includegraphics[width=\columnwidth]{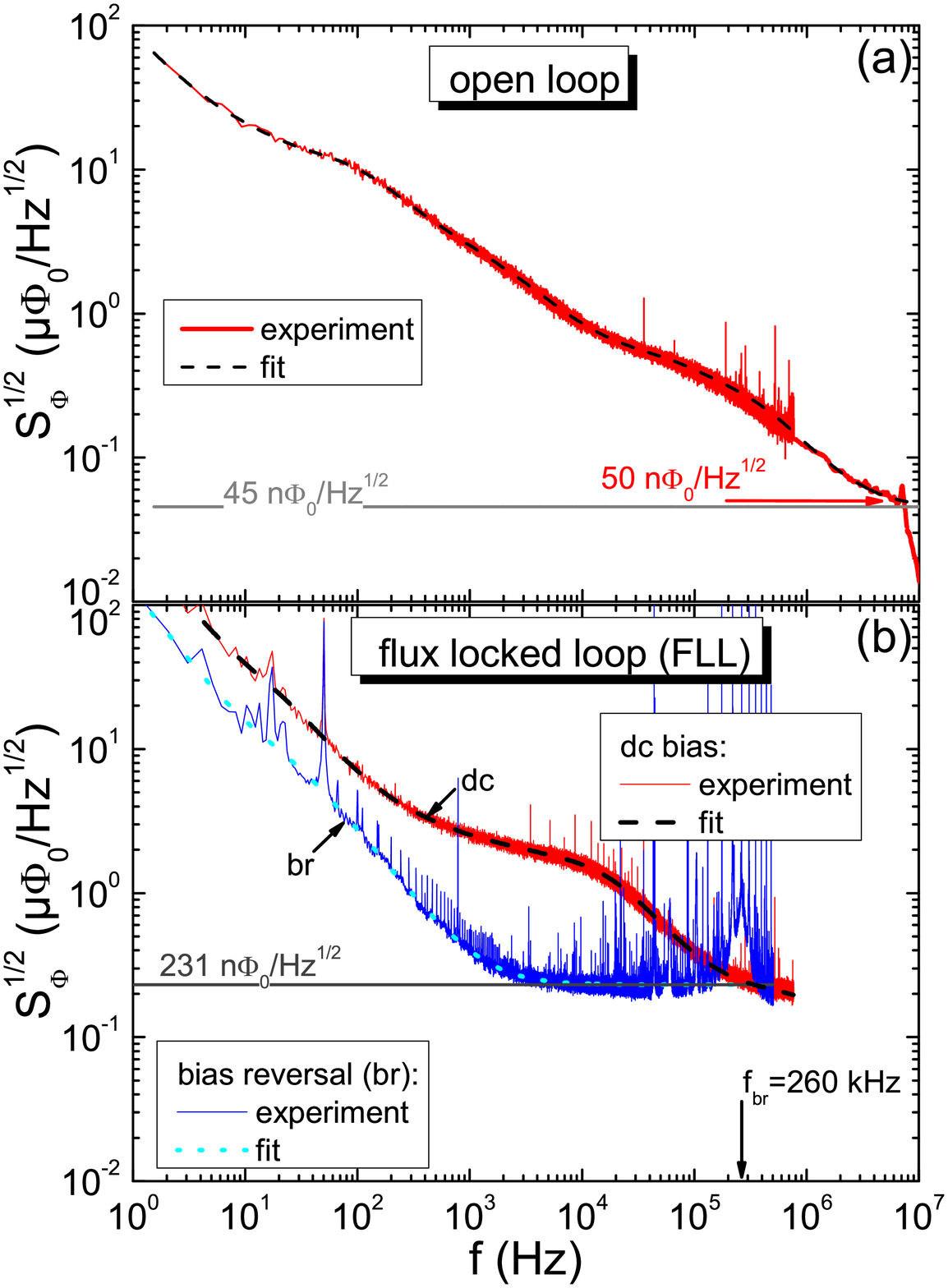}
\caption{ (Color online) Rms flux noise of SQUID-1.
%
% noise floor of the amplifier: 300 pV/Hz^1/2
%
(a) Measured in open loop mode at bias point 2 ($I=0.72\,$mA) in Fig.~\ref{Fig:dc-S1}(b).
Dashed line is a fit to the measured spectrum with white noise as indicated by horizontal line.
(b) Measured in FLL mode with dc bias and bias reversal ($|I|=0.43\,$mA, $V_\Phi=4.4\,{\rm mV}/\Phi_0$).
Vertical arrow indicates bias reversal frequency $f_\mathrm{br}$.
Dashed and dotted lines are fits to the spectra; horizontal lines indicate fitted white noise.}
\label{Fig:noise-S1}
\end{figure}

Figure \ref{Fig:noise-S1}(a) shows the rms spectral density of flux noise $S_\Phi^{1/2}(f)$ of SQUID-1, measured in open loop mode to reach the highest possible bandwidth of the readout electronics.
Due to the limitation in the maximum bias current of the readout electronics, noise spectra were taken at $I=0.72$\,mA with a transfer function $V_{\Phi}=4.5\,{\rm mV}/\Phi_0$ [cf.~point 2 in Fig.~\ref{Fig:dc-S1}(b)].
Up to the cutoff frequency $f_\mathrm{3dB}=7\,$MHz there is no white flux noise observable.
Instead, the flux noise scales roughly as $S_\Phi\propto 1/f$, with $S_\Phi^{1/2}\approx 10 \,\mu\Phi_0/{\rm Hz}^{1/2}$ at $f=100\,$Hz and $1\,\mu\Phi_0/{\rm Hz}^{1/2}$ at 10\,kHz.
This level of low-frequency excess noise is quite typical for YBCO GBJJ SQUIDs (also at $T=77\,$K) and has been ascribed to critical current fluctuations in the GBJJs \cite{Koelle99}.
However, due to the limitation by thermal white noise, typically between 1 and 10\,$\mu\Phi_0/{\rm Hz}^{1/2}$ for low-noise YBCO SQUIDs, this $f$-dependent excess noise has not been observed so far up to the MHz range.
We note that for YBCO nanoSQUIDs implementing cJJs \cite{Arpaia14}, a frequency-dependent 1/$f$-like excess noise at $T=8\,$K of almost the same level as for SQUID-1 was reported very recently, and was also attributed to critical current fluctuations.
For frequencies above 10\,kHz, the flux noise of the YBCO nanoSQUID in Ref.~[\onlinecite{Arpaia14}] was limited by amplifier background noise.

%%%%% Fig.3 %%%%%%%%%%%%%%%%%%%%%%%%%%%%%%%%%%%%%%%%%%%%%
%%%%% Fig.3 %%%%%%%%%%%%%%%%%%%%%%%%%%%%%%%%%%%%%%%%%%%%

For a more detailed analysis of the measured flux noise $S_\Phi(f)$, we applied an algorithm \cite{Sassier14} to decompose the noise spectra into a sum of Lorentzians $F_i(f)=F_{0,i}/[1+(f/f_{{\rm c,}i})^2]$ plus a white noise contribution $F_\mathrm{w}$.
The noise spectrum measured for SQUID-1 in open loop can be very well fitted by $F_\mathrm{op}(f)=F_\mathrm{w,op}+F_\mathrm{s,op}+\sum_{i=1}^{16}F_{{\rm op,}i}(f)$, i.e., the superposition of a white noise contribution with $F_\mathrm{w,op}^{1/2}=45\,{\rm n}\Phi_0/{\rm Hz}^{1/2}$ plus a $1/f^2$ spectrum $F_\mathrm{s,op}$ (i.e.~one or more Lorentzians with characteristic frequencies $f_\mathrm{c}$ well below 1\,Hz) with $F_\mathrm{s,op}^{1/2}(1\,{\rm Hz})=84\,\mu\Phi_0/{\rm Hz}^{1/2}$ plus 16 Lorentzians, with $f_{{\rm c,}i}$ ranging from 2.6\,Hz to 2.6\,MHz; for more details see Sec.~III of the Supplemental Material \cite{Suppl}.
Hence, the decomposition of the spectrum into Lorentzians yields an estimate of the white rms flux noise $S_\mathrm{\Phi,w}^{1/2}\approx 45\,{\rm n}\Phi _0/{\rm Hz}^{1/2}$ for SQUID-1.
We note that this value for $S_\mathrm{\Phi,w}^{1/2}$ is only a factor of 1.8 above the value which we obtain from numerical simulations of the coupled Langevin equations \cite{Chesca-SHB-2} at $T=4.2\,$K for the parameters of SQUID-1.

Taking the measured flux noise at 7\,MHz as an upper limit for $S_\mathrm{\Phi,w}^{1/2}$, we still obtain a very low white rms flux noise, i.e.~$S_\mathrm{\Phi,w}^{1/2}<50\,{\rm n}\Phi _0/{\rm Hz}^{1/2}$.
This more conservative estimate for the white rms flux noise level is an improvement by more than an order of magnitude compared to our non-optimized devices operated at 4.2\,K and compared to the lowest value reported so far for a YBCO SQUID (at 8\,K) very recently \cite{Arpaia14}.
Furthermore, this value is the same as the lowest value reported for a Pb SOT operated at 4.2\,K \cite{Vasyukov13} and among the lowest flux noise levels ever achieved for a SQUID \cite{VanHarlingen82,Awschalom88,Levenson-Falk13}.

For the geometry of SQUID-1, we calculate \cite{Woelbing14} a coupling factor $\phi_\mu = 13.4\,{\rm n}\Phi_0/\mu_\mathrm{B}$ (10\,nm above the YBCO film).
With $S_\mathrm{\Phi,w}^{1/2}<50\,{\rm n}\Phi _0/{\rm Hz}^{1/2}$, this yields an upper limit for the spin sensitivity (white noise limit) of $S_\mathrm{\mu,w}^{1/2}<3.7\,\mu_\mathrm{B}/{\rm Hz}^{1/2}$.
If we take the fitted white flux noise of $45\,{\rm n}\Phi _0/{\rm Hz}^{1/2}$, we obtain $S_\mathrm{\mu,w}^{1/2}=3.4\,\mu_\mathrm{B}/{\rm Hz}^{1/2}$.
Hence, the achieved performance matches very well the predictions of our recent optimization study \cite{Woelbing14}.

%%%%% Tab.1 %%%%%%%%%%%%%%%%%%%%%%%%%%%%%%%%%%%%%%%%%%%%%
\renewcommand{\arraystretch}{1.5}
\begin{table*}[t]
\caption{Parameters of optimized SQUID-1 and -2 and of SQUID-3 used for measurements on Fe nanowire.
Values for $V_\phi$ correspond to working points of noise measurements.
Values in brackets for $S_\mathrm{\Phi,w}^{1/2}$ and $S_\mathrm{\mu,w}^{1/2}$ of SQUID-1 are based on the fitted noise spectrum.
All devices have $d_\mathrm{Au}=70\,$nm.
SQUID-1 and -3 were measured at 4.2\,K, SQUID-2 was measured at 5.3\,K.}
\begin{center}
\tabcolsep1.5mm
\begin{tabular}{p{2cm} c c c c c c c c c c c c c c c}\hline\hline
        &$d$    & $l_{\rm c}$   & $l_{\rm J}$   & $w_{\rm c}$   & $w_{\rm J1}$  & $w_{\rm J2}$  & $\beta_L$ & $L$   & $I_{\rm c}$   & $R_{\rm N}$   & $I_{\rm c}R_{\rm N}$  & $V_\Phi$      & $S_\mathrm{\Phi,w}^{1/2}$ & $\phi_\mu$            & $S_\mathrm{\mu,w}^{1/2}$      \\
units   & nm    & nm            & nm            & nm            & nm            & nm            &           & pH    & $\mu$A        & $\Omega$      & mV                    & mV/$\Phi_0$   & n$\Phi_0/\rm{Hz}^{1/2}$   & n$\Phi_0/\mu_{\rm B}$ & $\mu_{\rm B}/\rm{Hz}^{1/2}$   \\\hline\hline
SQUID-1 & 120   & 190           & 350           & 85            & 210           & 160           & 1.8       & 3.9   & 960           & 2.0           & 1.92                  & 4.4           & $<50$ (45)                & 13                    & $<3.7$ (3.4)                  \\\hline
SQUID-2 & 120   & 230           & 370           & 100           & 180           & 230           & 0.94      & 6.3   & 311           & 2.5           & 0.78                  & 1.7           & $<83$                     & 12                    & $<6.7$                        \\\hline
SQUID-3 & 75    & 190           & 340           & 100           & 270           & 340           & 0.95      & 28    & 69            & 2.3           & 0.16                  & 0.65          &$<1450$                    & 15                    &$<98$                          \\\hline
\end{tabular}
\end{center}
\label{Tab:Parameters SQUID}
\end{table*}
%%%%% Tab.1 %%%%%%%%%%%%%%%%%%%%%%%%%%%%%%%%%%%%%%%%%%%%%

\subsubsection{FLL mode: dc bias vs bias reversal}
\label{subsubsec:FLL}

Although the achieved low level of white flux noise for SQUID-1 is encouraging, one certainly would like to extend such a low-noise performance down to much lower frequencies.
Therefore, we also performed noise measurements in FLL mode (with $\sim 700\,$kHz bandwidth) and compared measurements with dc bias and bias reversal (with $f_\mathrm{br}=260\,$kHz).
We note that the measurements in FLL mode were performed within a different cooling cycle, after SQUID-1 already showed a slight degradation in $I_\mathrm{c}$ \cite{Degradation}.
Still, we were able to find a working point (at $|I|=0.43\,$mA) which yielded almost the same transfer function, $4.4\,{\rm mV}/\Phi_0$, as for the measurement before degradation in open loop mode.

Figure \ref{Fig:noise-S1}(b) shows rms flux noise spectra taken with dc bias and bias reversal.
Comparing first the FLL dc bias measurement with the open loop data, we note that the noise levels at $f_\mathrm{br}$ coincide.
For $f<f_\mathrm{br}$ the noise levels of the open loop and FLL dc bias data are similar, however, the shape of the spectra differ, which we attribute to the above mentioned degradation and variations between different cooling cycles.
The dashed line in Fig.~\ref{Fig:noise-S1}(b) is a fit to the measured spectral density of flux noise by $F_\mathrm{dc}(f)=F_\mathrm{w,dc}+\sum_{i=1}^{15}F_{{\rm dc,}i}(f)$, i.e., the superposition of 15 Lorentzians, with $f_{{\rm c,}i}$ ranging from 0.8\,Hz to 6.8\,MHz, plus a white noise contribution $F_\mathrm{w,dc}^{1/2}=41\,{\rm n}\Phi_0/{\rm Hz}^{1/2}$, which we fixed to a value similar to the white noise level determined for the open loop measurement; for more details see Sec.~III of the Supplemental Material \cite{Suppl}.

Applying bias reversal, one expects a suppression of the contributions due to in-phase and out-of-phase critical current fluctuations of the GBJJs \cite{Koelle99}.
If the $f$-dependent excess noise below $f_\mathrm{br}$ would arise solely from $I_0$ fluctuations, one would expect in bias reversal mode a frequency-independent noise for frequencies below the peak at $f_\mathrm{br}$, at a level which is given by the noise measured at $f_\mathrm{br}$ in dc bias mode.
This is what we observe for frequencies down to a few kHz, with a $f$-independent noise $F_\mathrm{w,br}^{1/2}=231\,{\rm n}\Phi_0/{\rm Hz}^{1/2}$.
For lower frequencies, however, we still find a strong $f$-dependent excess noise in bias reversal mode, which hence cannot be attributed to $I_0$ fluctuations.

The spectral density of flux noise measured in bias reversal mode can be well approximated [cf.~dotted line in Fig.~\ref{Fig:noise-S1}(b)] by $F_\mathrm{br}(f)=F_\mathrm{w,br}+F_\mathrm{s,br}+\sum_{i=1}^{6}F_{{\rm br,}i}(f)$, with $F_\mathrm{s,br}^{1/2}(1\,{\rm Hz})=128\,\mu\Phi_0/{\rm Hz}^{1/2}$ and $f_{{\rm c,}i}$ of the six Lorentzians ranging from 21\,Hz to 5\,kHz; for more details see Sec.~III of the Supplemental Material \cite{Suppl}.

Obviously, below a few kHz the low-frequency excess noise is dominated by slow fluctuators, which cannot be attributed to $I_0$ fluctuations.
For different working points ($I$ and $I_\mathrm{mod}$) and also for other devices, the observation of low-$f$ excess noise in bias reversal mode was reproducible [cf.~flux noise data of SQUID-2 (from $T=6\,$K up to 65\,K) and of SQUID-3 (at $T=4.2\,$K) in Sec.~I and Sec.~II, respectively, of the Supplemental Material \cite{Suppl}].

Considering the narrow linewidths of the SQUID structures, we estimate a threshold field for trapping of Abrikosov vortices \cite{Stan04} to be well above 1\,mT.
Since the measurements were performed in magnetically shielded environment well below 100\,nT, the presence of Abrikosov vortices as the source of the observed low-$f$ fluctuators is very unlikely.

Low-frequency excess noise, which does neither arise from $I_0$ nor from vortex fluctuations, has been reported during the last decades for SQUIDs based on conventional superconductors like Nb, Pb, PbIn and Al, in particular at temperatures well below 1\,K \cite{Wellstood87}.
This issue has recently been revived due to the increasing interest in the development of flux qubits and SQUIDs for ultra-low temperature applications \cite{Drung11}.
Various models have been suggested to describe the origin of such low-$f$ excess noise, e.g. based on the coupling of magnetic moments associated with trapped electrons \cite{Koch07} or surface states \cite{deSousa07,Sendelbach08}, although the microscopic nature of defects as sources of excess 'spin noise' still remains unclear.

For YBCO SQUIDs, excess low-$f$ spin noise has not been addressed so far.
However, it seems quite likely that defects are also a source of magnetic fluctuators in SQUIDs based on cuprates or any other oxide superconductors.
Such defects could be present either in the thin film SQUID structures themselves, or in the substrates onto which the  thin films are grown, or at the interface between the thin film and the substrate.

The emergence and modification of magnetism at interfaces and surfaces of oxides, which are diamagnetic in the bulk, is currently an intensive field of research \cite{Venkatesan04,Pavlenko12,Coey13}.
For STO, oxygen vacancy-induced magnetism has been predicted \cite{Shein07}, and experimental studies suggest ferromagnetic ordering up to room temperature \cite{Khalid10}, e.g.~for defects induced by ion irradiation of single crystalline STO \cite{Potzger11}.
Furthermore, defect-induced magnetism in oxide grain boundaries and related defects have been suggested to be the intrinsic origin of ferromagnetism in oxides \cite{Straumal09}.

Obviously, further investigations on the impact and nature of such defects in our devices are needed and will be the subject of further studies.
Such studies will include detailed noise measurements (dc vs bias reversal, variable flux bias, temperature and magnetic field) to characterize and understand the $f$-dependent noise sources and, hopefully, eliminate them.
Furthermore, readout with bias reversal at higher frequency up to the MHz range in FLL mode has to be implemented, in order to maintain the achieved ultra-low white flux noise level down to lower frequencies.
And finally, for applications of our nanoSQUIDs, it will be important to avoid degradation in time.
This shall be achieved by adding a suitable passivation layer, however, without introducing $f$-dependent excess noise.

\section{SQUID-3: Magnetization reversal of Fe nanowire}
\label{sec:SQUID-3-FeNanowire}

As a proof of principle, we demonstrate nanoSQUID measurements on the magnetization reversal of a Fe nanowire which is encapsulated in a carbon nanotube (CNT) \cite{Leonhardt06}.
Such iron-filled CNTs (FeCNTs) are of fundamental interest with respect to studies on nanomagnetism.
Furthermore, they are attractive for various applications, e.g.~as tips in magnetic force microscopy \cite{Lipert10,Muehl12}.
The Fe-nanowire, which contains mainly single crystalline (ferromagnetic) $\alpha$-Fe, has a diameter $d_\mathrm{Fe}=39\,$nm and length $l_\mathrm{Fe}=13.8\,\mu$m.
The CNT has a diameter of $\sim 130\,$nm.
We note that this section is not directly related to the previous section in a sense to demonstrate the ultimate sensitivity of our devices on a magnetic nanoparticle with smallest still detectable signals and operation in strongest possible magnetic fields.
We rather want to show an example on the feasibility of using our YBCO nanoSQUIDs for practical applications.
As shown within this section, we can demonstrate signal-to-noise ratios which are clearly superior to micro-Hall measurements on similar nanowires.

%%%%% Fig.4 %%%%%%%%%%%%%%%%%%%%%%%%%%%%%%%%%%%%%%%%%%%%%
\begin{figure}[b]
\includegraphics[width=\columnwidth]{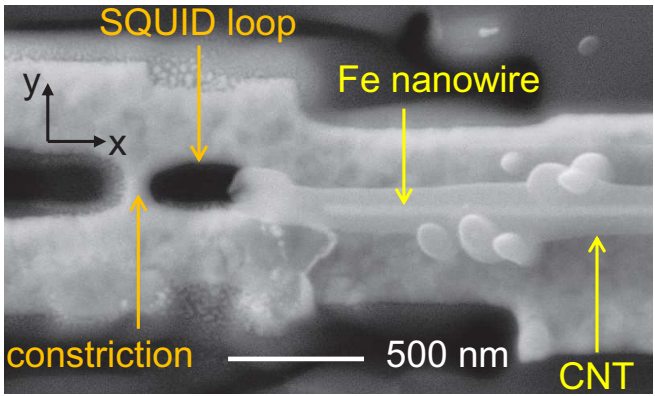}
\caption{(Color online) SEM image of SQUID-3 with Fe-wire filled carbon nanotube positioned close to the SQUID loop.}
\label{Fig:SEM-S3-FeCNT}
\end{figure}
%%%%% Fig.4 %%%%%%%%%%%%%%%%%%%%%%%%%%%%%%%%%%%%%%%%%%%%%

The FeCNT was positioned by a Kleindiek 3-axis manipulator inside a FIB-SEM combination onto SQUID-3, such that the distance between the left end of the Fe nanowire and the SQUID loop is $\sim 300\,$nm (cf.~Fig.~\ref{Fig:SEM-S3-FeCNT}).
We note that for optimum coupling of the stray field of the Fe nanowire into the SQUID, it is preferable to place the end of the Fe nanowire close to the edge of the SQUID loop opposite to the constriction.
At this location, the coupling factor is slightly smaller than directly on top of the constriction, however, it does not fall off very rapidly upon moving further away from the loop, as it is the case near the constriction \cite{Schwarz13}.
The Fe nanowire axis (its easy axis) was aligned as close as possible with the substrate plane [$(x,y)$ plane], with an inclination angle $\theta\,\approx 4\,^\circ$ and perpendicular to the grain boundary, which is oriented along the $y$-axis.
The inclination of the Fe wire axis with respect to the $x$-axis is $<1\,^\circ$.
The vertical distance (along the $z$-axis) between the nanowire axis (at its left end) and the surface of the YBCO film is $\approx 300\,$nm.

The measurements on the Fe nanowire were performed with the non-optimized SQUID-3.
This device has a significantly larger inductance (due to its smaller film thickness) and much smaller characteristic voltage, resulting in a much smaller transfer function $V_\Phi=0.65\,{\rm mV}/\Phi_0$, as compared to SQUID-1 and -2.
Magnetization reversal measurements on the FeCNT were performed with SQUID-3 operated in FLL dc bias mode up to $f=190\,$kHz.
At this frequency, the noise was limited by the readout electronics, which yields for SQUID-3 an upper limit of the white rms flux noise $S_\mathrm{\Phi,w}^{1/2} \le 1.45\,\mu\Phi_0/{\rm Hz}^{1/2}$.
Below $\sim 40\,$kHz, SQUID-3 showed $f$-dependent excess noise with $S_\Phi^{1/2}\approx 8\,\mu\Phi_0/{\rm Hz}^{1/2}$ at $f=100\,$Hz and $S_\Phi^{1/2}\approx 20\,\mu\Phi_0/{\rm Hz}^{1/2}$ at $f=10\,$Hz, with an approximately $1/f^2$ increase of $S_\Phi$ below 10\,Hz.
Some experimentally determined parameters of SQUID-3 are listed in Tab.~\ref{Tab:Parameters SQUID}.
Details on low-field electric transport and noise characteristics of SQUID-3 are presented in Sec.~II of the Supplemental Material \cite{Suppl}.

For magnetization reversal measurements of the Fe nanowire on top of SQUID-3, the sample was mounted in a high-field setup, which allows to apply magnetic fields up to $\mu_0 H=7\,$T \cite{Schwarz13}.
To minimize coupling of the external magnetic field $H$ into the SQUID, the SQUID loop (in the $(x,y)$ plane) has been aligned parallel to the field.
To minimize coupling of the external field into the GBJJs, the grain boundary (along the $y$-axis) was aligned perpendicular to the applied field.
The alignment of the SQUID with respect to the applied field direction was performed by an Attocube system including two goniometers with perpendicular tilt axes and one rotator.
In this configuration, the external field $H$ is applied along the $x$-axis (cf.~Fig.~\ref{Fig:SEM-S3-FeCNT}), and the angle between $H$ and the Fe nanowire axis is given by $\theta$.

%%%%%% Fig.5 %%%%%%%%%%%%%%%%%%%%%%%%%%%%%%%%%%%%%%%%%%%%
\begin{figure}[t]
\includegraphics[width=\columnwidth]{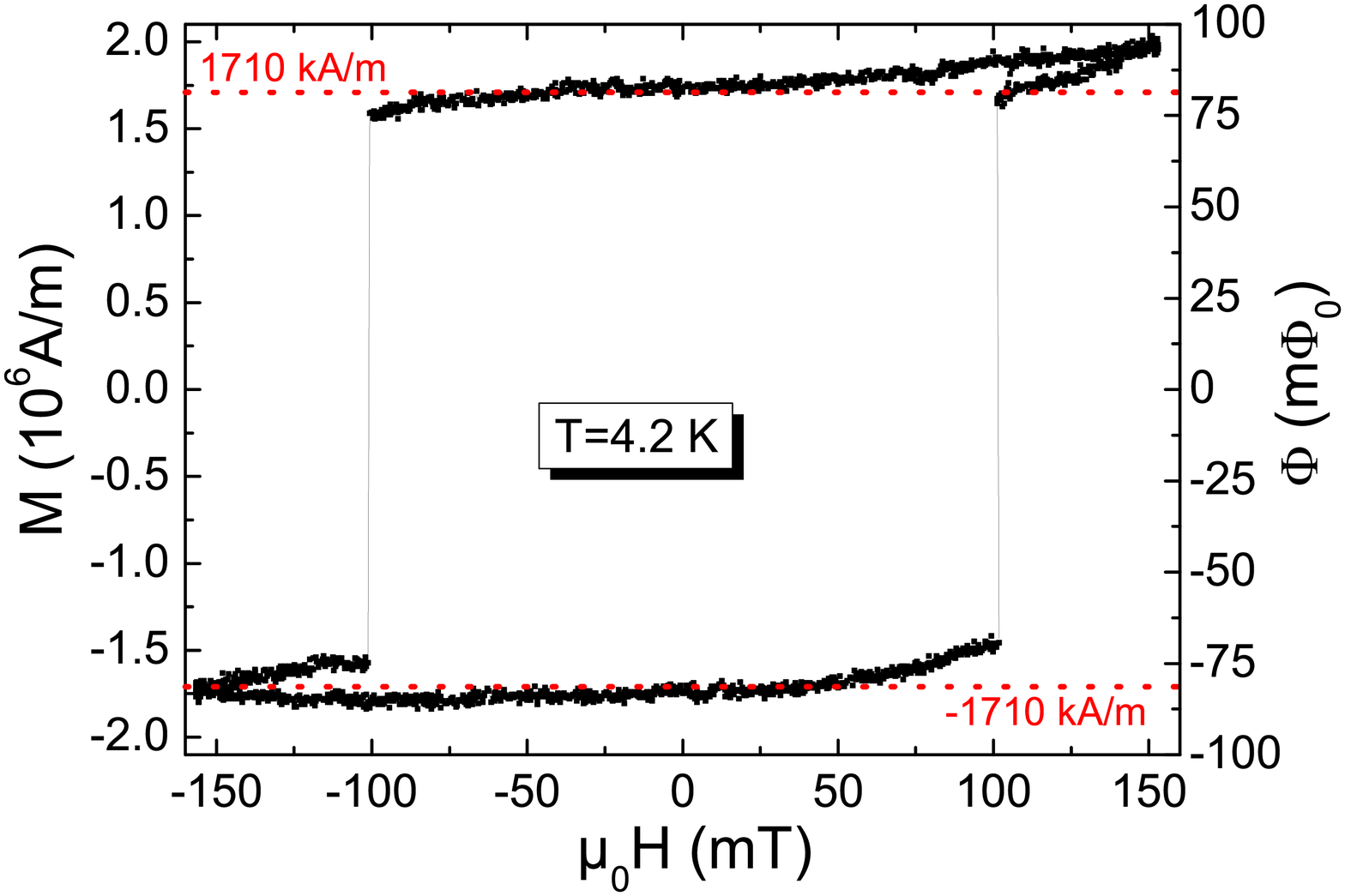}
\caption{(Color online)
Hysteresis loop $\Phi(H)$ of the Fe-nanowire detected with SQUID-3 (operated in FLL dc bias mode with cutoff frequency $\sim 190\,$kHz, at optimum working point with $V_\Phi=0.65\,{\rm mV}/\Phi_0$).
Switching of the magnetization occurs at $\pm\mu_0H_\mathrm{n} = \pm 101\,$mT.
The residual field $\mu_0H_\mathrm{res}=4.0\,$mT was subtracted.
Left axis indicates corresponding magnetization $M=\Phi/\phi_M$; the dashed lines indicate the literature value of the saturation magnetization $\pm M_\mathrm{s}$.}
\label{Fig:Hysterese}
\end{figure}
%%%%% Fig.5 %%%%%%%%%%%%%%%%%%%%%%%%%%%%%%%%%%%%%%%%%%%%%

Figure \ref{Fig:Hysterese} shows the flux signal $\Phi(H)$ detected by SQUID-3, while sweeping $H$, at a rate $\mu_0\partial H/\partial t\approx1\,$mT/s.
At the fields $\pm\mu_0 H_\mathrm{n}= \pm 101\,$mT, abrupt changes by $\Delta\Phi \approx 150\,{\rm m}\Phi_0$ clearly indicate magnetization reversal of the Fe nanowire.
The shape of the $\Phi(H)$ curve indicates magnetization reversal of a single domain particle.
The slope of the curve in the interval $-H_\mathrm{n}\le H\le H_\mathrm{n}$ depends strongly on the alignment of the SQUID with respect to the applied field.
Hence, this slope can be attributed, at least partially, to the coupling of the external field to the SQUID loop.
The hysteresis in the signals for $|H|\,\gapprox\, 100\,$mT is typically observed also for our SQUIDs measured in the high-field setup without MNPs coupled to them.
Hence, this hysteresis is attributed to a spurious magnetization signal from our setup or from the above mentioned magnetic defects close to the nanoSQUID, rather than being generated by the nanowire.

In order to convert from magnetic flux detected by the SQUID to magnetization of the Fe nanowire, we follow the approach described in Ref.~[\onlinecite{Nagel13}].
We numerically calculate the coupling factor $\phi_\mu(\hat{e}_\mu,\vec{r}_\mathrm{p})$ for a point-like MNP with orientation $\hat{e}_\mu$ of its magnetic moment at position $\vec{r}_\mathrm{p}$ in the 3D space above the SQUID \cite{Woelbing14}.
These simulations take explicitly into account the geometry of SQUID-3 and are based on London theory \cite{Khapaev03}.
We then assume that the Fe nanowire is in its fully saturated state, with saturation magnetization $M_\mathrm{s}$, with all moments oriented along the wire axis.
The corresponding saturation flux coupled to the SQUID is denoted as $\Phi_\mathrm{s}$.
The ratio $\Phi_\mathrm{s}/M_\mathrm{s}$ is obtained by integration of the coupling factor $\phi_\mu$ over the volume $V_\mathrm{Fe}$ of the Fe wire, at its given position, determined from SEM images.
This yields
%
%%%%%%%%%%%%%%%%% eq.1 %%%%%%%%%%%%%%%%%%%%%%%%%%%%%%%%%
\begin{equation}
\phi_M \equiv \frac{\Phi_\mathrm{s}}{M_\mathrm{s}} = \int_{V_\mathrm{Fe}}\,\phi_\mu(\vec{r_\mathrm{p}})\,dV = 47.6\,\frac{{\rm n}\Phi_0}{{\rm A m}^{-1}}\;.
\label{eq:int}
\end{equation}
%%%%%%%%%%%%%%%%% eq.1 %%%%%%%%%%%%%%%%%%%%%%%%%%%%%%%%%
%
From this we calculate $\Phi_\mathrm{s}=M_\mathrm{s}\phi_M= 81.4\,{\rm m}\Phi_0$, with $M_\mathrm{s}=1710\,$kA/m taken from literature \cite{Schabes91}.
The comparison with the measured flux signals $\pm 82.5\,{\rm m}\Phi_0$ at $H=0$ shows very good agreement.
The left axis in Fig.~\ref{Fig:Hysterese} shows the magnetization axis, scaled as $M=\Phi/\phi_M$, with the horizontal dotted lines indicating the literature value $M_\mathrm{s}=\pm 1710\,$kA/m.
Hence, the measured flux signals are also in quantitative agreement with the assumption that the Fe nanowire switches to a fully saturated single domain state.

In Ref.~[\onlinecite{Lipert10}] it was shown for a similar FeCNT that the nucleation field $H_\mathrm{n}$ changes with $\theta$ in a way which is typical for nucleation of magnetization reversal via the curling mode \cite{Aharoni66} in ferromagnetic nanowires as opposed to uniform rotation of the magnetic moments in small enough MNPs as described by the Stoner-Wolfarth model \cite{Skomski08}.
For switching via curling mode one obtains for $\theta=0$ the simple relation $H_\mathrm{n}=M_\mathrm{s}a/2$, with a negligible increase well below 1\,\% with $\theta=4\,^\circ$ \cite{Aharoni97}.
Here, $a=1.08\,(2\lambda_\mathrm{ex}/d_\mathrm{Fe})^2$, with the exchange length $\lambda_\mathrm{ex}=\sqrt{4\pi A/(\mu_0 M_\mathrm{s}^2)}$ and the exchange constant $A$ \cite{Schabes91}.
For $d_\mathrm{Fe}=39\,$nm and with $\lambda_\mathrm{ex}=5.8\,$nm \cite{Schabes91}, we obtain $a=0.0955$, and with $M_\mathrm{s}=1710\,$kA/m we obtain an estimate of the nucleation field $H_\mathrm{n}=103\,$mT, which is in very good agreement with the experimentally observed value.

Finally, we note that the SQUID measurement yields a noise amplitude of $\sim 1 {\rm m}\Phi_0$, which is two orders of magnitude smaller than the detected signal upon magnetization reversal.
For comparison, measurements on a similar Fe nanowire by micro-Hall magnetometry yielded a noise amplitude which was about one order of magnitude below the switching signal \cite{Lipert10}.
This means that the use of our nanoSQUID improves the signal-to-noise ratio by about one order of magnitude.

\section{Conclusions}
\label{sec:Conclusions}

In conclusion, we fabricated and investigated optimized YBCO nanoSQUIDs based on grain boundary Josephson junctions.
For our best device, an upper limit for the white flux noise level $S_{\Phi}^{1/2} < 50\,{\rm n}\Phi_0/{\rm Hz}^{1/2}$ in magnetically shielded environment could be determined, which corresponds to a spin sensitivity $S_\mu^{1/2} \equiv S_\Phi^{1/2}/\phi_\mu = 3.7\,\mu_\mathrm{B}/{\rm Hz}^{1/2}$ for a magnetic nanoparticle located 10\,nm above the constriction in the SQUID loop.
Here, the coupling factor $\phi_\mu$ was determined by numerical simulations based on London theory, which takes the device geometry into account.
An obvious drawback of YBCO grain boundary junction nanoSQUIDs is the frequency-dependent excess noise, which extends up to the MHz range for optimized devices with ultra-low flux noise in the white noise limit.
To eliminate $1/f$ noise, a bias reversal scheme was applied, which only partially reduced the frequency-dependent excess noise.
Hence, in addition to critical current fluctuations, spin noise which is possibly due to fluctuations of defect-induced magnetic moments in the SrTiO$_3$ substrate is a major issue, which has to be studied in more detail for further improvement of the nanoSQUID performance at low frequencies.
Nevertheless, we demonstrated the suitability of the YBCO nanoSQUIDs as detectors for magnetic nanoparticles in moderate magnetic fields by measuring the magnetization reversal of an iron nanowire that was placed close to the SQUID loop.
Switching of the magnetization was detected at $\mu_0H \approx \pm 100\,$mT, which is in very good agreement with nucleation of magnetization reversal via curling mode.

\acknowledgments

T.~Schwarz acknowledges support by the Carl-Zeiss-Stiftung.
M.~J.~Mart\'{i}nez-P\'{e}rez acknowledges support by the Alexander von Humboldt Foundation.
We gratefully acknowledge fruitful discussions with D.~Drung (PTB Berlin) and technical support by M.~Turad and R.~L\"offler (instrument scienists of the core facility LISA$^+$).
This work was supported by the Nachwuchswissenschaftlerprogramm of the Universit\"{a}t T\"{u}bingen, by the Deutsche Forschungsgemeinschaft (DFG) via projects KO 1303/13-1, MU 1794/3-2 and SFB/TRR 21 C2 and by the EU-FP6-COST Action MP1201.


\begin{thebibliography}{65}
\bibitem [{Deg()}]{Degradation}%
  \BibitemOpen
  \href@noop {} {}\bibinfo {note} {From measurements on a similar device we
  found an almost linear decrease in $I_c$ with time, with a $\sim 20\,$\%
  reduction in $I_c$ after 40 days. This rate of degradation is typical for
  most of our devices. A possible explanation of this effect is outdiffusion of
  oxygen from the submicron GBJJs along the $a-b$ plane of the YBCO thin
  film.}\BibitemShut {Stop}%
\end{thebibliography}
\end{document}